\documentclass[sigconf]{acmart}

\AtBeginDocument{%
  }

\copyrightyear{2026}
\acmYear{2026}
\setcopyright{cc}
\setcctype{by}
\acmConference[SIGIR '26]{Proceedings of the 49th International ACM SIGIR Conference on Research and Development in Information Retrieval}{July 20--24, 2026}{Melbourne, VIC, Australia}
\acmBooktitle{Proceedings of the 49th International ACM SIGIR Conference on Research and Development in Information Retrieval (SIGIR '26), July 20--24, 2026, Melbourne, VIC, Australia}
\acmDOI{10.1145/3805712.3808645}
\acmISBN{979-8-4007-2599-9/2026/07}

\begin{document}

\title{Verifiable User Simulation for Search and Recommendation Systems}

\author{Chenglong Ma}
\authornote{Chenglong is the main contact of this tutorial.}
\email{chenglong.ma@rmit.edu.au}
\orcid{0000-0002-6745-4029}
\affiliation{%
  \institution{RMIT University}
  \city{Melbourne}
  \state{Victoria}
  \country{Australia}
}

\author{Xinye Wanyan}
\email{xinye.wanyan@student.rmit.edu.au}
\orcid{0009-0002-7264-1803}
\affiliation{%
  \institution{RMIT University}
  \city{Melbourne}
  \state{VIC}
  \country{Australia}
}

\author{Danula Hettiachchi}
\email{danula.hettiachchi@rmit.edu.au}
\orcid{0000-0003-3875-5727}
\affiliation{%
  \institution{RMIT University}
  \city{Melbourne}
  \state{VIC}
  \country{Australia}
}

\author{Ziqi Xu}
\email{ziqi.xu@rmit.edu.au}
\orcid{0000-0003-1748-5801}
\affiliation{%
  \institution{RMIT University}
  \city{Melbourne}
  \state{VIC}
  \country{Australia}
}

\author{Yongli Ren}
\email{yongli.ren@rmit.edu.au}
\orcid{0000-0002-3137-9653}
\affiliation{%
  \institution{RMIT University}
  \city{Melbourne}
  \state{VIC}
  \country{Australia}
}

\author{Jeffrey Chan}
\email{jeffrey.chan@rmit.edu.au}
\orcid{0000-0002-7865-072X}
\affiliation{%
  \institution{RMIT University}
  \city{Melbourne}
  \state{VIC}
  \country{Australia}
}

\renewcommand{\shortauthors}{Chenglong Ma et al.}

\begin{abstract}
Large-language-model (LLM) based user simulation is increasingly adopted for evaluating search engines, recommender systems, and retrieval-augmented generation pipelines, yet most simulators remain opaque: it is difficult to determine \emph{why} a simulated user made a particular choice or whether that choice is consistent with the intended user profile. Compounding this, recent research shows that LLMs can produce biased or discriminatory responses depending on user background characteristics such as language, education level, and cultural context, raising concerns about the equitable treatment of minority and disadvantaged groups.  This half-day, in-person tutorial introduces a proposed design-and-audit framework that treats a user simulator as a \emph{verifiable engineering artefact} composed of seven auditable components---structured Persona, task-aware Contract, matched human-vs-agent Execution, auditable Trace, persona-aligned Verification, structured Feedback, and a Refinement loop that updates personas and contracts.  Through two hands-on mini-labs on recommendation-list evaluation and search-query formulation, participants will inspect simulator behaviour end-to-end, distinguish diagnostic discrepancy analysis from statistical validation, and apply checks for fidelity, credibility, and demographic bias. The tutorial targets information retrieval and recommender systems researchers and practitioners interested in user behaviour simulation and responsible AI.
\end{abstract}



\maketitle

\section{Tutorial Overview}

This half-day tutorial presents a proposed framework for building and auditing LLM-based
user simulators for search and recommendation systems.  It
combines three concise lecture segments with two hands-on mini-labs in which
participants use a browser-based environment to construct personas, specify
task contracts, compare human and agent behaviour, inspect traces, and refine
simulator specifications.  All presenters will attend and deliver the tutorial
on-site.  The tutorial includes short Q\&A and setup checkpoints before the
hands-on activities.

The intended audience is intermediate: information retrieval and recommender systems researchers,
practitioners, and graduate students interested in LLM-based user simulation,
evaluation methodology, fairness auditing, or human-centred AI.  Participants
should be familiar with core information retrieval or recommender systems concepts such as queries, relevance,
ranking, search engine result pages, and collaborative or content-based
filtering.  No prior experience with LLM prompting, agent design, or
simulation frameworks is assumed.

\paragraph{Prior delivery.}
An earlier 75-minute version was delivered at the ARC Centre of Excellence for Automated Decision-Making \& Society (ADM+S) Summer School \footnote{See \url{https://www.admscentre.org.au/event/2026-adms-summer-school/}.}
in Melbourne in 2026 to approximately 50 participants from information
retrieval, media, and social science.  Feedback from that session motivated a
deeper treatment of verification methodology and additional hands-on work.  The
SIGIR version expands the session into a half-day tutorial by adding a full
treatment of the seven-component verification framework, two structured
mini-labs, and explicit coverage of fairness verification and
demographic-bias detection~\cite{gallegos2024bias}.

\section{Motivation}

User simulation has a long history in information retrieval evaluation, including
pre-LLM simulation of interactive search behaviour~\cite{maxwell2016Simiir}.
LLMs have changed the design space by making it easier to instantiate
persona-conditioned agents for offline evaluation of search engines,
recommender systems, and retrieval-augmented generation
pipelines~\cite{gao2024llmsurvey,zhang2024agent4rec}.  Recent studies show
that LLMs can produce behavioural patterns resembling human
samples~\cite{argyle2023out} and sustain coherent personas across extended
interactions~\cite{park2023generative}.  These capabilities have accelerated
their use for relevance assessment, query generation, click modelling,
conversational evaluation, and other settings where collecting real user data is
costly, slow, or ethically constrained.

The apparent fidelity of simulated behaviour can be misleading.  Current
simulators are often opaque: researchers cannot easily determine why an agent
made a choice, whether that choice follows the intended user profile, or how
sensitive the outcome is to prompts, model selection, and decoding parameters.
At the same time, evidence shows that LLM responses can vary with language,
education level, culture, and other user background characteristics, producing
biased or discriminatory outcomes~\cite{gallegos2024bias,tao2024cultural,das2024unveiling}.
In search and recommendation, such variation can distort evaluation results
and lead to unfair treatment of minority or disadvantaged groups.  Without
traceable evidence and verification checks, simulation-driven evaluation rests
on assumptions that are difficult to reproduce or audit.

We therefore treat a user simulator as a \emph{verifiable engineering
artefact}: a system whose behaviour can be inspected, attributed, and
systematically improved.  We present the seven-component decomposition below as
a practical design and audit framework.
Its purpose in the tutorial is to give
participants a vocabulary and workflow for making simulation assumptions
explicit.
\begin{itemize}\setlength{\itemsep}{0pt}
  \item \textbf{Persona:} structured, editable user attributes, including
        demographics, preferences, knowledge level, and interaction style.
  \item \textbf{Contract:} task-specific constraints and preferences that scope the
        simulation without overfitting to a single scenario.
  \item \textbf{Execution:} matched human and persona-conditioned agent runs
        on the same task, enabling behavioural comparison.
  \item \textbf{Trace:} auditable records of inputs, actions, rationales, timestamps,
        model identifiers, and provider configurations.
  \item \textbf{Verification:} persona-aligned checks for fidelity, credibility,
        fairness, and the source of observed discrepancies, rather than accuracy
        alone.
  \item \textbf{Feedback:} attributable evidence identifying where a
        persona, contract, or agent output failed.
  \item \textbf{Refinement:} an update loop that revises personas and contracts based
        on verification outcomes.
\end{itemize}

The framework is grounded in two domains.  In recommendation-list evaluation,
participants act as an assigned persona, choose items from a fixed list, and
compare their selections with an LLM agent's trace.  In search-query
formulation, participants generate queries from a backstory and verify
agent-generated queries against the same persona specification.  A single
human--agent comparison is not presented as statistical validation; it is used as
a diagnostic probe.  Participants label discrepancies as likely agent deviation,
persona ambiguity, contract underspecification, role-player interpretation, or
model stochasticity, and then discuss what evidence would be needed for stronger
empirical claims.  Both tasks run in a browser-based environment and require no
external API access during the session.

\section{Learning Objectives}

By the end of the tutorial, participants will be able to:
\begin{enumerate}\setlength{\itemsep}{0pt}
  \item define structured personas and task-aware contracts for simulation
        studies;
  \item design matched human and agent execution protocols for controlled
        behavioural comparison;
  \item collect and inspect execution traces that support reproducibility and
        post-hoc analysis;
  \item apply verification checks for persona fidelity, rationale credibility,
        fairness, and discrepancy attribution;
  \item identify cases where simulated behaviour varies systematically with
        language, culture, education level, or other background attributes;
        and
  \item use structured feedback to refine personas and contracts across
        simulation rounds while separating pedagogical inspection from
        population-level validation.
\end{enumerate}

\section{Relevance to SIGIR}

User simulation has long been part of SIGIR's evaluation methodology,
including simulated interaction for search evaluation~\cite{maxwell2016Simiir}
and recent workshops and tutorials on LLM-based evaluation for search and
recommendation~\cite{siro2025llm4eval,balog2025tutorial}.  The rapid adoption
of LLM agents has increased the need for practical methods that make simulated
behaviour reproducible, interpretable, and fair.  Several presenters have also
worked on LLM-based user simulation for recommender system evaluation, including
personality-driven and temporal-aware simulators~\cite{ma2022nest,ma2025pub,wanyan2025temporal}.

This tutorial complements prior work that surveys simulation techniques or
focuses on specific application domains~\cite{balog2025tutorial, balog2024tutorial}.
Its focus is verification: how to inspect simulator outputs, attribute
behaviour to persona and contract choices, and decide whether the resulting
evidence is credible enough to support an evaluation claim.  The tutorial also
places fairness auditing inside the core simulation workflow, reflecting
growing evidence that LLM-based simulators can amplify demographic biases when
left unchecked~\cite{gallegos2024bias,tao2024cultural}.

The material is relevant to researchers designing simulation-based evaluation,
practitioners building LLM-driven search or recommendation systems, and
reviewers assessing the credibility of simulation studies.  It aligns with
SIGIR's broader interest in reproducible evaluation~\cite{ferro2017reproducibility}
and responsible deployment of generative models in information access systems.

\section{Format and Schedule}

The tutorial is an in-person half-day session of 180 minutes.  It interleaves
short lectures with hands-on work so that each conceptual block is followed by
practice on a concrete simulation task.  Short question and setup checkpoints are
included before each mini-lab.

\begin{itemize}\setlength{\itemsep}{2pt}
  \item \textbf{0--5 min: Welcome and orientation.}  Tutorial goals,
        logistics, and overview of the seven-component framework.
  \item \textbf{5--20 min: From user simulation to Persona and Contract design.}
        Brief pre-LLM context, structured persona attributes, persona
        libraries, task-aware contracts, and examples from search and
        recommendation.
  \item \textbf{20--25 min: Questions and setup check.}  Clarify concepts and
        ensure participants can access the web environment.
  \item \textbf{25--40 min: Execution and traces.}  Matched human--agent
        runs, trace-schema design, and recording of actions,
        rationales, timestamps, and model configurations.
  \item \textbf{40--45 min: Questions and task walkthrough.}  Resolve setup
        issues and preview the first mini-lab workflow.
  \item \textbf{45--90 min: Mini-Lab 1, recommendation.}  Participants act as
        an assigned persona, select items from a fixed recommendation list,
        compare their selections with an LLM agent, inspect the trace, classify
        discrepancies, and complete a short debrief.
  \item \textbf{90--105 min: Break and informal questions.}
  \item \textbf{105--120 min: Verification and refinement.}  Fidelity,
        credibility, fairness, discrepancy attribution, evidence linked
        feedback, and the persona/contract update loop.
  \item \textbf{120--125 min: Questions and setup check.}  Clarify the
        verification rubric before the second mini-lab.
  \item \textbf{125--165 min: Mini-Lab 2, search.}  Participants formulate
        queries from an assigned backstory, compare them with agent-generated
        queries, apply verification checks, and draft structured feedback.
        An optional extension uses a fixed corpus and BM25 result page for
        result selection and query reformulation.
  \item \textbf{165--180 min: Wrap-up and discussion.}  Synthesis, open Q\&A,
        and structured participant feedback.
\end{itemize}

The session includes 85 minutes of mini-lab work, 15 minutes of closing
discussion, and 15 minutes of explicit question/setup checkpoints.  This design
keeps more than half of the tutorial participant-active while reducing the risk
that browser or task-access issues consume lab time.

\section{Materials and Participation}

All materials will be prepared in advance and distributed at the start of the
session.  The package includes a slide deck, a hands-on workbook, reusable
Persona and Contract JSON schemas, a Trace schema, verification rubrics,
feedback forms, and an annotated reading list.  The workbook provides
step-by-step instructions for both mini-labs and is designed to remain useful
as a template for participants' own evaluation workflows.  The verification
rubric asks participants to record the evidence for each judgement and to
separate agent errors from ambiguous personas, underspecified contracts, human
role-playing variance, and model stochasticity.

The hands-on activities use a browser-based demo environment at
\url{https://pslab.simubox.org}.  The environment provides persona and
contract editors, recommendation lists, search backstories, pre-generated LLM
agent outputs, trace viewers, evaluation results, feedback views, and refined
persona editing.  Participants need only a laptop or tablet with web access;
all computation is server-side.  A downloadable offline backup will be
available for connectivity failures.

After the tutorial, slides, workbooks, schemas, and a recorded summary video
will be released at \url{https://sigir2026tutorial.simubox.org}.  This release
is intended to support reuse by researchers conducting simulation-based
evaluation or adapting the framework to their own IR and RecSys pipelines.

\paragraph{Accessibility and ethics.}
Slides and workbooks will be reviewed for accessible contrast, font size, and
screen-reader compatibility.  Tutorial activities use synthetic personas by
default and can be completed without submitting identifiable information.  Any
optional collection of anonymised interaction traces for improving the platform
will be subject to RMIT Human Research Ethics review and the safeguards
required by that process.  Fairness discussions will focus on trace-based
auditing, measurement choices, and feedback-loop controls rather than
individual-level labelling.

\section{Presenter Expertise}

The presenter team combines expertise in information retrieval, recommender
systems, human-centred AI, responsible AI, and empirical evaluation.  The team
has published related work on personality-driven and temporal-aware LLM user
simulation for recommender system evaluation~\cite{ma2022nest,ma2025pub,wanyan2025temporal}
and has prior experience delivering the material in an interactive summer-school
setting.







\paragraph{Chenglong Ma} Chenglong Ma is a Research Fellow at the RMIT University node of the ADM+S Centre, specialising in user simulation and evaluation methods for recommendation and search systems.
His work focuses on verifiable, persona-conditioned simulation and
fairness-aware evaluation.
He has experience delivering interactive research sessions, including a
75-minute workshop at the ADM+S Centre Summer School on verifiable user
simulation. He also has 4 years of teaching experience in practical data science.

\paragraph{Xinye Wanyan} Xinye Wanyan is a PhD student at the RMIT University node of the ADM+S Centre, where her PhD research focuses on user behaviour simulation for the evaluation of recommender systems. Her current work introduces a temporal-aware mechanism that enables agents to model evolving user behaviour based on historical interactions. She also designs automated user-profile generation methods to bridge the gap between interaction sequences and agent personas, improving the realism and effectiveness of simulated users.


\paragraph{Danula Hettiachchi} Danula Hettiachchi is a Lecturer in the School of Computing Technologies at RMIT University and an Associate Investigator at ADM+S. His research spans crowdsourcing, social computing, HCI, and responsible AI, with a focus on designing and analysing human studies and auditing human--AI decision processes. He brings experience in delivering and supporting interactive teaching and community-facing tutorial activities.


\paragraph{Ziqi Xu} Ziqi Xu is a Lecturer in the School of Computing Technologies at RMIT University. His research focuses on responsible AI, including fairness, causal inference, and explainable machine learning, with recent work on evaluation methodologies and LLM-based user behaviour simulation for recommendation and information retrieval. He teaches across data science and machine learning subjects, supporting hands-on tutorial delivery.


\paragraph{Yongli Ren} Yongli Ren is an Associate Professor in the School of Computing Technologies at RMIT University. His research focuses on trustworthy and equitable recommender systems, including bias detection and mitigation, transparency, and explanation, as well as simulation-based evaluation with controllable LLM user simulators beyond accuracy-only metrics. He has extensive teaching experience and has received multiple teaching awards at RMIT.

\paragraph{Jeffrey Chan} Jeffrey Chan is a Professor in the School of Computing Technologies at RMIT University. His research interests lie in machine learning, recommender systems, responsible AI, data-driven optimisation and interdisciplinary research that combines these fields to solve social and industry-based applications. He has worked with industry and non-profit partners in sustainability, manufacturing and health.

\begin{acks}
This work was conducted at the ARC Centre of Excellence for Automated Decision-Making and Society (ADM+S) and funded by the Australian Research Council (CE200100005). It was supported by computing resources from RACE (RMIT Advanced Cloud Ecosystem) and RMIT's Enabling Impact Platforms. This research was undertaken on the unceded lands of the Woi wurrung and Boon wurrung language groups of the eastern Kulin Nation. We pay our respects to Elders past and present, and to Aboriginal and Torres Strait Islander peoples and their continuing connections to land, sea, sky, and community.
\end{acks}

\bibliographystyle{ACM-Reference-Format}
\balance
\bibliography{refs}

\end{document}